\documentclass[a4paper]{article}
\usepackage{amscd,amsmath,amssymb}

\newcommand{\bQ}{{\overline Q}{}}

\newcommand{\be}{\begin{equation}}
\newcommand{\ee}{\end{equation}}
\newcommand{\bea}{\begin{eqnarray}}
\newcommand{\eea}{\end{eqnarray}}

\newcommand{\ba}{\begin{array}} \newcommand{\ea}{\end{array}}

\def\im{{\rm i}}

\newcommand{\nn}{\nonumber}

\begin{document}

\begin{center}
{\huge\bf $N=4$ $\ell$-conformal Galilei superalgebras  }\\
\vspace{0.5cm}
{\huge\bf inspired by $D(2,1;\alpha)$ supermultiplets  }
\end{center}
\vspace{1cm}

\begin{center}
{\Large\bf Anton Galajinsky${}^{a}$  and  Sergey Krivonos${}^{b}$
}
\end{center}

\vspace{0.2cm}

\begin{center}
${}^a$ {\it
Tomsk Polytechnic University,
634050 Tomsk, Lenin Ave. 30, Russia} \\

\vspace{0.3cm}

${}^b$ {\it
Bogoliubov  Laboratory of Theoretical Physics, JINR,
141980 Dubna, Russia}

\vspace{0.5cm}

{\tt galajin@tpu.ru, krivonos@theor.jinr.ru}
\end{center}
\vspace{2cm}

\begin{abstract}\noindent
$N=4$ supersymmetric extensions of the $\ell$--conformal Galilei algebra are constructed by properly extending the Lie superalgebra associated with the most general $N=4$ superconformal group in one dimension $D(2,1;\alpha)$. If the acceleration generators in the superalgebra form analogues of the irreducible $(1,4,3)$--, $(2,4,2)$--, $(3,4,1)$--, and $(4,4,0)$--supermultiplets of $D(2,1;\alpha)$, the parameter $\alpha$ turns out to be constrained by Jacobi identities. In contrast, if the tower of the acceleration generators resembles a component decomposition of a generic real superfield, which is a reducible representation of $D(2,1;\alpha)$, $\alpha$ remains arbitrary.
An $N=4$ $\ell$--conformal Galilei superalgebra recently proposed in [Phys. Lett. B 771 (2017) 401] is shown to be a particular instance of a more general construction in this work.
\end{abstract}


\vskip 0.5cm


\newpage

\setcounter{equation}{0}
\section{Introduction}

Current exploration of the non--relativistic version of the AdS/CFT--correspondence brings into focus nonrelativistic superconformal algebras. They are usually constructed by
choosing a proper subalgebra in a relativistic superconformal algebra (see, e.g., the discussion in \cite{SY}), or by implementing to the latter a nonrelativistic contraction possibly
accompanied by proper projections \cite{HU1}--\cite{MS}. An alternative and more straightforward possibility is to start with one or another representation of nonrelativistic conformal algebra, introduce extra fermionic degrees of freedom, and add supersymmetry charges by hand. Other generators are then unambiguously fixed from the requirement of the closure of the full superalgebra (see, e.g., \cite{PH}--\cite{GM} and references therein).

Apart from holographic applications, supersymmetric extensions of nonrelativistic conformal algebras are of interest in many--body quantum mechanics and near horizon black hole physics.
Following the proposal in \cite{GT}, according to which $n\to\infty$ limit of the $n$--particle $N=4$ superconformal Calogero model may provide a microscopic description of
the near horizon extreme Reissner–-Nordstr\"om black hole, various $N=4$ superconformal many--body models in one dimension have been constructed and investigated (see, e.g., \cite{GLP}--\cite{HV} and references therein). Higher--dimensional generalizations of the studies in \cite{GLP}--\cite{HV}  necessarily invoke supersymmetric extensions of nonrelativistic conformal algebras. Though $N=2$ models have been constructed in \cite{AG}, the instance of $N=4$, which is believed to the maximum value for which the construction of interacting models is feasible, remains completely unexplored. Worth mentioning also is the recent studies of superintegrable systems associated with the angular sector of a generic conformal mechanics \cite{HKLN,HLNS}. The models with $N=4$ conformal Galilei supersymmetry might be of interest in that context as well.

Conformal extension of the Galilei algebra involves a (half)integer parameter $\ell$ \cite{NOR,Henkel}. It can be written in a remarkably succinct way
\bea
&& \left[ L_m, L_n \right] = -\im\, (m-n) L_{n+m},\; \qquad \left[ L_n, {\bf U}_k\right] = -\im\, \left( \ell\, n -k\right) {\bf U}_{n+k}, \;
\nonumber
\eea
where $n,m = -1,0,1$ and $k=-\ell,\ldots,\ell$. The leftmost relation implies that $L_n$ form the conformal algebra in one dimension $so(2,1)$, $(L_{-1},L_{0},L_{1})$ being the generators of time translation, dilatation, and special conformal transformation, respectively, while the rightmost equation means that ${\bf U}_k$ has the conformal weight $\ell$. Above we omitted the conventional $so(d)$--rotations which also enter the algebra. In particular ${\bf U}_k$ carry an extra vector index of $so(d)$. Here and in what follows we suppress it and use boldface letters to designate generators which transform as vectors under the action of $so(d)$. A link to the acceleration generators $C^{(n)}_A$ in \cite{NOR}, where $n=0,\dots,2\ell$ and $A=1,\dots,d$ is a vector index of $so(d)$, is provided by the identification ${\bf U}_{m}=\im\, C^{(m+\ell)}_A$. In particular, ${\bf U}_{-\ell}$ and ${\bf U}_{-\ell+1}$ generate spatial translations, and Galilei boosts, while higher values of $m$ correspond to constants accelerations. Because the structure relations of the conformal extension of the Galilei algebra above involve a (half)integer number $\ell$, it is customary to call it the $\ell$--conformal Galilei algebra \cite{NOR}.

In a very recent work \cite{GM}, an $N=4$ supersymmetric extension of the $\ell$--conformal Galilei algebra was constructed by combining the generators of spatial symmetries from the $\ell$--conformal Galilei algebra and those underlying the most general superconformal group in one dimension $D(2,1;\alpha)$. The value of the group parameter $\alpha$ was fixed from the requirement that the resulting superalgebra was finite--dimensional. The analysis revealed $\alpha=-\frac 12$ thus reducing $D(2,1;\alpha)$ to $OSp(4|2)$. Note that the instance of an $N=4$, $\ell=1$ conformal Galilei superalgebra in three and four spatial dimensions was previously studied in \cite{FL}. The superalgebra was obtained by implementing the In\"on\"u--Wigner contraction to the relativistic superconformal algebra. However, similarly to the derivation of the $\ell$--conformal Galilei algebra itself,
the effectiveness of the contraction method beyond $\ell=1$ seems dubious.

In order to close the full superalgebra, a chain of extra generators was introduced in \cite{GM} which were interpreted as bosonic and fermionic partners of the acceleration generators. Their realization in terms of differential operators in superspace very much resembled a component decomposition of a generic real superfield.\footnote{For a similar consideration of $N=2$ $\ell$--conformal Galilei superalgebras based on real and chiral supermultiplets see \cite{IM,AKT}. Note that in Appendix of \cite{AKT} an attempt was made to construct an N=4 supersymmetric extension of the $\ell$--conformal Galilei algebra. However, the analysis relied upon a specific $D$--module representation of $D(2,1;\alpha)$ which led the authors to conclude that an infinite number of generators were needed in order to close the superalgebra. Note also that in \cite{AKT}
the conformal weights $(\ell,\ell-1/2)$ have been assigned to the acceleration generators. Our analysis below suggests that an alternative possibility involving $(\ell, \ell+1/2)$ is feasible which would result in yet another version of an $N=2$ $\ell$--conformal Galilei superalgebra.}. Because a real superfield provides a reducible representation of $D(2,1;\alpha)$, it is natural to wonder whether $N=4$ supersymmetric extensions of the $\ell$--conformal Galilei algebra exist in which the acceleration generators and their bosonic and fermionic partners form an analog of an irreducible representation of $D(2,1;\alpha)$. The goal of this work is to provide an affirmative answer to this question and to construct novel $N=4$ supersymmetric extensions of the $\ell$--conformal Galilei algebra.

As is known, irreducible supermultiples of $D(2,1;\alpha)$ are classified according to the number of physical bosonic and fermionic degrees of freedom as well as the number of auxiliary bosons \cite{IKL1}.\footnote{In the conventional notation the symbol $(b,f,a)$ designates a supermultiplet which contains $b$ physical bosons, $f$ physical fermions,
and $a=f-b$ auxiliary bosons \cite{IKL1}.} In Sect. 2 we discuss in detail $N=4$ supersymmetric extensions of the $\ell$--conformal Galilei algebra which rely upon
the $(1,4,3)$--, and $(3,4,1)$--supermultiplets of $D(2,1;\alpha)$. Similar construction involving the $(4,4,0)$--supermultiplet is represented in Sect. 3. Sect. 4 is focused on the
$(2,4,2)$--supermultiplet in which case $D(2,1;\alpha)$ should be reduced to $SU(1,1|2)\times U(1)$, where $U(1)$ stands for the central charge \cite{IKL1}. In the first three cases the group parameters $\alpha$ and $\ell$
turn out to be related by Jacobi identities. In Sect. 5 a reducible supermultiplet associated with a generic real scalar superfield is analyzed. An $N=4$ supersymmetric extensions of the $\ell$--conformal Galilei algebra
is constructed for which $\alpha$ is arbitrary. The superalgebra recently proposed in \cite{GM} is shown to be a particular instance of that in Sect. 5. Realizations in terms of differential operators in superspace are discussed in Sect. 6. We summarize our results and discuss possible further developments in the concluding Sect. 7.

\section{Acceleration generators vs $(1,4,3)$--, $(3,4,1)$--multiplets of $D(2,1;\alpha)$}

Our strategy for constructing $N=4$ supersymmetric extensions of the $\ell$--conformal Galilei algebra is to combine the generators of accelerations and those of the Lie superalgebra associated with $D(2,1;\alpha)$. In the succinct notation adopted in this work the structure relations of the latter read \cite{IKL}
\bea\label{Dalg}
&& \left[ L_m, L_n \right] = -\im\, (m-n) L_{n+m},\; \left[ V^{ab}, V^{cd} \right] = -\im\, \left( \epsilon^{ac} V^{bd} +\epsilon^{bd} V^{ac}\right),\;
\left[ W^{ij}, W^{kl} \right] = -\im\, \left( \epsilon^{ik} W^{jl} +\epsilon^{jl} W^{ik}\right), \nn \\
&& \left[ L_m, Q_r^{ai}\right]=-\im\, \left(\frac{m}{2}-r\right) Q_{r+m}^{ai}, \;
\left[ V^{ab},Q_r^{ci}\right] =-\frac{\im}{2} \left( \epsilon^{ac} Q_r^{bi}+\epsilon^{bc} Q_r^{ai}\right), \;
\left[ W^{ij},Q_r^{ak}\right] =-\frac{\im}{2} \left( \epsilon^{ik} Q_r^{aj}+\epsilon^{jk} Q_r^{ai}\right),\nn \\
&&\left\{Q_r^{ai},Q_s^{bj}\right\} = -2\left( \epsilon^{ab}\epsilon^{ij}L_{r+s}+\alpha (r-s) \delta_{r+s,0}\epsilon^{ab} W^{ij}-(1+\alpha)
(r-s) \delta_{r+s,0}\epsilon^{ij} V^{ab}\right).
\eea
Bosonic generators include $L_n$, $n=-1,0,1$, which span $so(2,1)$--subalgebra, along with $V^{ab}=V^{ba}$, $a,b = 1,2$, and $W^{ij}=W^{ji}$, $i,j =1,2$, which form two commuting $su(2)$--subalgebras. The fermionic generators $Q_r^{ai}$ carry spinors indices with respect to both the $su(2)$--subalgebras and an extra index $r= -\frac{1}{2}, \frac{1}{2}$ which separates (complex conjugate) supersymmetry charges from their superconformal partners.\footnote{Denoting the supersymmetry charges and their superconformal partners by $Q^i$ and $S^i$, respectively, the link to the notation in \cite{GM} is provided by the relations
$Q_{-\frac 12}^{1 i}=\bar Q^i$, $Q_{-\frac 12}^{2 i}=Q^i$, $Q_{\frac 12}^{1 i}=-\bar S^i$, $Q_{\frac 12}^{2 i}=-S^i$, where the bar designates complex conjugation. Note that in \cite{GM} $su(2)$--spinor indices were denoted by small Greek letters.} As is seen from (\ref{Dalg}), $(L_n,Q_r^{ai},V^{ab},W^{ij})$ have the conformal weights $(1,\frac 12,0,0)$, respectively.

As the next step, let us introduce $2\ell+1$ acceleration generators ${\bf U}_m$, $m=-\ell,\ldots,\ell$, which have the conformal weight $\ell$ and are inert under both the $su(2)$--transformations from $D(2,1;\alpha)$. For general reasons, the bracket of $Q_r^{ai}$ and ${\bf U}_m$ should yield a fermionic superpartner, which we denote by ${\bf S}_r^{ai}$, while the structure relation involving
$Q_r^{ai}$ and ${\bf S}_r^{ai}$ must produce a bosonic superpartner, say ${\bf A}_m^{ij}={\bf A}_m^{ji}$.\footnote{One could equally well consider the case when the bosonic partner carries two spinor indices with respect to $su(2)$ generated by $V^{ab}$.} Suppressing the lower index for a moment, one reveals the triplet $({\bf U},{\bf S}^{ai},{\bf A}^{ij})$ which looks analogous to the irreducible $(1,4,3)$--supermultiplet of $D(2,1;\alpha)$ \cite{IKL1}. In contrast to the analysis in \cite{GM}, in this section we choose to terminate further proliferation of extra bosonic and fermionic partners of ${\bf U}_m$ and consider a superalgebra in which the full set of the acceleration generators form an analog of the $(1,4,3)$--supermultiplet of $D(2,1;\alpha)$.

Turning to precise structure relations, two options are available. One can either assign the conformal weights $(\ell, \ell+\frac 12, \ell+1)$ to the triplet $({\bf U}_n,{\bf S}_r^{ai},{\bf A}_m^{ij})$, or, alternatively, choose the descending sequence $(\ell, \ell-\frac 12, \ell-1)$. To put it in other words, one can identify ${\bf U}_n$ with the lowest/highest component of the supermultiplet.

In the former case, the non--vanishing brackets among $({\bf U}_n,{\bf S}_r^{ai},{\bf A}_m^{ij})$ and the $D(1,2;\alpha)$--generators read
\bea\label{bosbos}
&& \left[ L_n, {\bf U}_m\right] = -\im\, \left( \ell\, n -m\right) {\bf U}_{n+m}, \; \left[ L_n, {\bf S}_r^{ai}\right] = -\im \, \left( \left(\ell+\frac{1}{2}\right)n -r\right){\bf S}_{n+r}^{ai},\nn \\
&&
\left[ L_n, {\bf A}_m^{ij}\right] = -\im\, \left( \left(\ell+1\right) n -m\right) {\bf A}_{n+m}^{ij}, \; \left[ W^{ij},{\bf A}_n^{kl}\right] =-\im\, \left( \epsilon^{ik} {\bf A}_n^{jl}+\epsilon^{jl} {\bf A}_n^{ik}\right),
\nn \\
&& \left[ V^{ab},{\bf S}_r^{ci}\right] =-\frac{\im}{2} \left( \epsilon^{ac} {\bf S}_r^{bi}+\epsilon^{bc} {\bf S}_r^{ai}\right), \;
\left[ W^{ij},{\bf S}_r^{ak}\right] =-\frac{\im}{2} \left( \epsilon^{ik} {\bf S}_r^{aj}+\epsilon^{jk} {\bf S}_r^{ai}\right),\nn \\
&&
\left[ W^{ij},{\bf A}_n^{kl}\right] =-\im\, \left( \epsilon^{ik} {\bf A}_n^{jl}+\epsilon^{jl} {\bf A}_n^{ik}\right), \; \left[ Q_r^{ai}, {\bf U}_m\right] = \im {\bf S}_{r+m}^{ai}, \nn \\
&&
\left[ Q_r^{ai}, {\bf A}_m^{jk} \right] = \im\, \left( 2\left(\ell+1\right) r - m\right) \left( \epsilon^{ij} {\bf S}_{r+m}^{ak}+\epsilon^{ik} {\bf S}_{r+m}^{aj}\right), \;\nn \\
&&
\left\{ Q_r^{ai}, {\bf S}_s^{bj}\right\} =\left( \left(2 \ell +1\right) r-s \right)\epsilon^{ab}\epsilon^{ij} {\bf U}_{r+s} -\epsilon^{ab} {\bf A}_{r+s}^{ij},
\eea
where the range of values for the indices labeling different members of the set $({\bf U}_n,{\bf S}_r^{ai},{\bf A}_m^{ij})$ is determined by their conformal weights
\bea\label{range}
&&
{\bf U}_n: n=-\ell,\ldots,\ell, \; \qquad  {\bf S}_r^{ai}: r= -\ell-\frac{1}{2},\ldots,\ell+\frac{1}{2}, \; \qquad {\bf A}_{m}^{ij}: m = -\ell-1, \ldots, \ell+1.
\eea
Note that all the factors in (\ref{bosbos}) which explicitly involve the parameter $\ell$ are designed so as to keep the full superalgebra finite dimensional. They balance in a proper way the index range in the previous formula and the appearance of the acceleration generators on the left and right hand sides of the structure relations (\ref{bosbos}).
Finally, one can verify that the only non--trivial Jacobi identity involves $(S_r^{ai},Q_s^{bj},Q_q^{ck})$ which links $\alpha$ to $\ell$
\be\label{alpha1}
\alpha = \ell +1.
\ee

Before we proceed to the second option mentioned above, a word of caution is needed. Though the analogy with the $(1,4,3)$--supermultiplet of $D(2,1;\alpha)$ proved rather helpful in constructing (\ref{bosbos}), strictly speaking, the triplets $({\bf U}_n,{\bf S}_r^{ai},{\bf A}_m^{ij})$ fail to produce a set of the conventional $(1,4,3)$--supermultiplets. This is because the ranges of values of the indices $n$, $r$ and $m$ in (\ref{range}) are different. To put it in other words, there is a mismatch
in the number of bosonic and fermionic components available in (\ref{range}) and that needed to furnish a set of the conventional $(1,4,3)$--supermultiplets.

If the conformal weights $(\ell, \ell-\frac 12, \ell-1)$ are assigned to the triplet $({\bf U}_n,{\bf S}_r^{ai},{\bf A}_m^{ij})$, the structure relations (\ref{bosbos}) and the index range (\ref{range}) are modified accordingly
\bea\label{bosbos2}
&& \left[ L_n, {\bf U}_m\right] = -\im\, \left( \ell\, n -m\right) {\bf U}_{n+m}, \; \left[ L_n, {\bf S}_r^{ai}\right] = -\im \, \left( \left(\ell-\frac{1}{2}\right)n -r\right) {\bf S}_{n+r}^{ai},\nn \\
&&\left[ L_n, {\bf A}_m^{ij}\right] = -\im\, \left( \left(\ell-1\right) n -m\right) {\bf A}_{n+m}^{ij}, \; \left[ Q_r^{ai}, {\bf U}_m\right] = \im \left( 2 \ell\, r -m\right) {\bf S}_{r+m}^{ai}, \nn \\
&& \left[ Q_r^{ai}, {\bf A}_m^{jk} \right] = \im\, \left( \epsilon^{ij} {\bf S}_{r+m}^{ak}+\epsilon^{ik} {\bf S}_{r+m}^{aj}\right), \;
\left\{ Q_r^{ai}, {\bf S}_s^{bj}\right\} =\epsilon^{ab}\epsilon^{ij} {\bf U}_{r+s} -\left( \left( 2\ell -1\right)r -s\right) \epsilon^{ab} {\bf A}_{r+s}^{ij}.
\eea
where
\bea
{\bf U}_n: n=-\ell,\ldots,\ell, \; \qquad {\bf S}_r^{ai}: r= -\ell+\frac{1}{2},\ldots,\ell-\frac{1}{2}, \; \qquad  {\bf A}_{m}^{ij}: m = -\ell+1, \ldots, \ell-1
\eea
Here and in what follows we omit brackets among the acceleration generators and $V^{ab}$, $W^{ij}$. They have the standard form of $su(2)$ transformations and are constructed by analogy with (\ref{Dalg}) and (\ref{bosbos}).
Like above, the only nontrivial Jacobi identity which needs to be verified involves $(S_r^{ai},Q_s^{bj},Q_q^{ck})$. It relates the group parameters $\alpha$ and $\ell$
\be\label{alpha2}
\alpha = - \ell.
\ee
As compared to the previous case, the superalgebra (\ref{bosbos2}) is defined for $\ell\geq 1$. Note that ${\bf S}_r^{ai}$ has two fewer components than its analog in (\ref{range}), while for ${\bf A}_m^{ij}$ there is a decrease of four components.

Concluding this section, it is worth mentioning that a seemingly different possibility of constructing $N=4$ supersymmetric extensions of the $\ell$--conformal Galilei algebra arises if one interchanges conformal weights assigned to ${\bf U}_m$ and ${\bf A}^{ij}_m$ and treats one of the components of ${\bf A}^{ij}_m$ as the generator of accelerations in the original $l$--conformal Galilei algebra.\footnote{In the particular case of three spatial dimensions, one can discard an extra $so(3)$ vector index attached to ${\bf A}^{ij}_m$ and identify one of the internal $su(2)$--subalgebras generated by $W^{ij}$ with the spatial rotation symmetry. For $\ell=1$ such a consideration was discussed in \cite{FL}.} This produces mirror versions for the superalgebras (\ref{bosbos}) and (\ref{bosbos2}). In particular, the triplet $({\bf A}_m^{ij},{\bf S}_r^{ai},{\bf U}_n)$ may be loosely linked to the irreducible $(3,4,1)$--supermultiplet of $D(2,1;\alpha)$. However, such superalgebras are not independent. It turns out that the formal change $(\ell +1) \to \tilde\ell$ in the mirror version of (\ref{bosbos}) yields (\ref{bosbos2}), while the interchange $(\ell -1) \to \tilde\ell$ in the mirror version of (\ref{bosbos2}) results in (\ref{bosbos}).

\section{Acceleration generators vs $(4,4,0)$--multiplet of $D(2,1;\alpha)$}

The construction above can be readily generalized to the case when the acceleration generators in an $N=4$ $\ell$--conformal Galilei superalgebra form an analog of the $(4,4,0)$--supermultiplet of $D(2,1;\alpha)$.
It suffices to introduce the bosonic generators ${\bf q}_n^{i A}$, with $n=-\ell,\ldots,\ell$, $i=1,2$, $A=1,2$, and their fermionic partners ${\bf S}_r^{a A}$, for which $r= -\ell-\frac{1}{2},\ldots,\ell+\frac{1}{2}$, $a=1,2$, $A=1,2$, and to impose the structure relations
\bea\label{bosbosh}
&& \left[ L_n, {\bf q}_m^{i A}\right] = -\im\, \left( \ell\, n -m\right) {\bf q}_{n+m}^{i A}, \;
\left[ L_n, {\bf S}_r^{a A}\right] = -\im \, \left( \left(\ell+\frac{1}{2}\right)n -r\right) {\bf S}_{n+r}^{a A},\nn \\
&& \left[ Q_r^{ai}, {\bf q}_m^{j A}\right] = \im \epsilon^{i j} {\bf S}_{r+m}^{a A}, \; \left\{ Q_r^{a i}, {\bf S}_s^{b A}\right\} =-2 \left( \left(2 \ell +1\right) r-s \right)\epsilon^{ab} {\bf q}_{r+s}^{i A},
\eea
along with the constraint required by the Jacobi identities $\alpha = - 2 (\ell+1)$. Here we omitted the standard brackets with $W^{ij}$ and $V^{ab}$. Suffice it to say that ${\bf q}_m^{i A}$ carries a spinor index $i$ with respect to the $su(2)$--subalgebra generated by $W^{ij}$, while ${\bf S}_r^{a A}$ transforms as a spinor of the $su(2)$--subalgebra generated by $V^{ab}$. There is also an external $su(2)$--automorphism which acts upon the index $A$.

An alternative possibility is to assign the conformal weights $\ell$ and $\ell-\frac{1}{2}$ to ${\bf q}_n^{i A}$ and ${\bf S}_r^{a A}$, respectively, where $n=-\ell,\ldots,\ell$, $r= -\ell+\frac{1}{2},\ldots,\ell-\frac{1}{2}$. This yields the superalgebra
\bea\label{bosbosh2}
&& \left[ L_n, {\bf q}_m^{i A}\right] = -\im\, \left( \ell\, n -m\right) {\bf q}_{n+m}^{i A}, \;
\left[ L_n, {\bf S}_r^{a A}\right] = -\im \, \left( \left(\ell-\frac{1}{2}\right)n -r\right){\bf S}_{n+r}^{a A},\nn \\
&&
\left[ Q_r^{ai}, {\bf q}_m^{j A}\right] = -2 \im \left(2 \ell r -m\right)  \epsilon^{i j} {\bf S}_{r+m}^{a A}, \; \left\{ Q_r^{a i}, {\bf S}_s^{b A}\right\} =  \epsilon^{ab} {\bf q}_{r+s}^{i A},
\eea
for which the Jacobi identities constrain $\alpha$ as follows: $\alpha = 2  \ell$. The superalgebra is defined for $\ell\geq \frac 12$ and ${\bf S}_r^{a A}$ has two fewer components than its analog in (\ref{bosbosh}).
Note that in the particular case of four spatial dimensions, one can discard an extra $so(4)$ vector index attached to ${\bf q}_m^{i A}$ and identify the internal symmetry $su(2)\oplus su(2)$ with the spatial rotation invariance $so(4) \simeq su(2)\oplus su(2)$. For $d=4$ and $\ell=1$ an alternative consideration based upon the In\"on\"u--Wigner contraction of the relativistic superconformal algebra was presented in \cite{FL}.

\section{Acceleration generators vs $(2,4,2)$--multiplet of $SU(1,1|2)$}

As is known, a rigorous description of a supermultiplet of the type $(2,4,2)$ is attained at $\alpha=-1$, which reduces $D(2,1;\alpha)$ to
$SU(1,1|2)$ \cite{IKL1}. In the succinct notation adopted in this work the structure relations of $su(1,1|2)$ read
\bea\label{su112}
&& \left[ L_m, L_n \right] = -\im\, (m-n) L_{n+m},\; \left[ V^{ab}, V^{cd} \right] = -\im\, \left( \epsilon^{ac} V^{bd} +\epsilon^{bd} V^{ac}\right),\nn \\
&& \left[ L_m, Q_r^{a}\right]=-\im\, \left(\frac{m}{2}-r\right) Q_{r+m}^{a}, \;
\left[ V^{ab},Q_r^{c}\right] =-\frac{\im}{2} \left( \epsilon^{ac} Q_r^{b}+\epsilon^{bc} Q_r^{a}\right), \nn \\
&& \left[ L_m, \bQ_r^{a}\right]=-\im\, \left(\frac{m}{2}-r\right) \bQ_{r+m}^{a}, \;
\left[ V^{ab},\bQ_r^{c}\right] =-\frac{\im}{2} \left( \epsilon^{ac} \bQ_r^{b}+\epsilon^{bc} \bQ_r^{a}\right), \nn \\
&&\left\{Q_r^{a},\bQ_s^{b}\right\} = -2\left( \epsilon^{ab} L_{r+s} - (r-s) V^{ab} +(r-s)\epsilon^{ab} J\right),
\eea
where $J$ is the central charge operator.

In order to construct an $N=4$ $\ell$--conformal Galilei superalgebra associated with the $(2,4,2)$--supermultiplet of $SU(1,1|2)$, let us introduce
the (complex conjugate) bosonic generators $({\bf q}_n,{\bar {\bf q}}_n)$ and $({\bf p}_m,{\bar {\bf p}}_m)$, which have the conformal weights
$\ell$ and $\ell+1$, respectively, and their (complex conjugate) fermionic partners $({\bf S}_r^{a}, {\bar {\bf S}}_r^{a})$, $a=1,2$, with the conformal weight $\ell+1/2$, and impose the structure relations
\bea\label{bosbosChiral}
&& \left[ L_n, {\bf q}_m\right] = -\im\, \left( \ell\, n -m\right) {\bf q}_{n+m}, \;
 \left[ L_n, {\bar {\bf q}}_m\right] = -\im\, \left( \ell\, n -m\right){\bar {\bf q}}_{n+m}, \nn \\
&& \left[ L_n, {\bf S}_r^{a}\right] = -\im \, \left( \left(\ell+\frac{1}{2}\right)n -r\right) {\bf S}_{n+r}^{a},\;
\left[ L_n, {\bar {\bf S}}_r^{a}\right] = -\im \, \left( \left(\ell+\frac{1}{2}\right)n -r\right){\bar {\bf S}}_{n+r}^{a},\nn \\
&& \left[ L_n, {\bf p}_m\right] = -\im\, \left( \left(\ell+1\right)\, n -m\right) {\bf p}_{n+m}, \;
\left[ L_n, {\bar {\bf p}}_m\right] = -\im\, \left( \left(\ell+1\right)\, n -m\right){\bar {\bf p}}_{n+m} , \nn \\
&&
\left[ V^{ab},{\bf S}_r^{c}\right] =-\frac{\im}{2} \left( \epsilon^{ac} {\bf S}_r^{b}+\epsilon^{bc} {\bf S}_r^{a}\right),\; \left[ V^{ab},{\bar {\bf S}}_r^{c}\right] =-\frac{\im}{2} \left( \epsilon^{ac} {\bar {\bf S}}_r^{b}+\epsilon^{bc} {\bar {\bf S}}_r^{a}\right),\nn \\
&& \left[ J, {\bf q}_m\right] = \im (\ell+1) {\bf q}_m,\; \left[ J, {\bar {\bf q}}_m\right] = -\im (\ell+1) {\bar {\bf q}}_m,\; \left[ J, {\bf p}_m\right] = \im (\ell+1) {\bf p}_m,\; \left[ J, {\bar {\bf p}}_m\right] = -\im (\ell+1) {\bar {\bf p}}_m, \nn \\
&& \left[ J,{\bf S}_r^{a} \right] = \im (\ell+1) {\bf S}_r^a, \; \left[ J, {\bar {\bf S}}_r^{a} \right] = - \im (\ell+1){\bar {\bf S}}_r^a, \; \left[Q_r^a, {\bf q}_n\right] = -2 \im {\bf S}_{r+n}^a, \;
\left[\bQ_r^a, {\bar {\bf q}}_n\right] = -2 \im {\bar {\bf S}}_{r+n}^a, \nn \\
&& \left\{ Q_r^a, {\bf S}_s^b\right\}= \epsilon^{ab} {\bf p}_{r+s}, \; \left\{ \bQ_r^a, {\bf S}_s^b\right\}= \left( ( 2\ell+1) r -s\right) \epsilon^{ab} {\bf q}_{r+s}, \;
\left\{ Q_r^a, {\bar {\bf S}}_s^b\right\}= -\left( ( 2\ell+1) r -s\right) \epsilon^{ab} {\bar {\bf q}}_{r+s}, \nn \\
&& \left\{ \bQ_r^a, {\bar {\bf S}}_s^b\right\}= -\epsilon^{ab} {\bar {\bf p}}_{r+s}, \;
\left[ Q_r^a, {\bar {\bf p}}_n\right]=2 \im \left( 2 (\ell+1)r-n\right) {\bar {\bf S}}_{r+n}^a, \;
\left[ \bQ_r^a,{\bf p}_n\right]=2 \im \left( 2 (\ell+1)r-n\right) {\bf S}_{r+n}^a.
\eea
As usual, the index range is fixed by the conformal weights and the requirement that the superalgebra is finite dimensional
\be
({\bf q}_n, {\bar {\bf q}}_n): n=-\ell,\ldots,\ell, \; ({\bf S}_r^{a}, {\bar {\bf S}}_r^a): r= -\ell-\frac{1}{2},\ldots,\ell+\frac{1}{2}, \;
({\bf p}_n, {\bar {\bf p}}_n): n=-\ell-1,\ldots,\ell+1.
\ee
It is straightforward to verify that the Jacobi identities do hold for (\ref{bosbosChiral}). In this framework either ${\bf q}_n+{\bar {\bf q}}_n$ or $\im ({\bf q}_n-{\bar {\bf q}}_n)$ can be identified with the acceleration generators in the original $\ell$--conformal Galilei algebra. Note that, though originally $J$ enters (\ref{su112}) as a central charge, in the full superalgebra (\ref{su112}), (\ref{bosbosChiral}) its status is changed to assign the $U(1)$--charge to the acceleration generators.

\section{Acceleration generators vs reducible multiplet of $D(2,1;\alpha)$}

One can also consider a more general situation when the acceleration generators in an $N=4$ $\ell$--conformal Galilei superalgebra form an analog of a general unconstrained real superfield of $D(2,1;\alpha)$. Let us introduce the bosonic generators ${\bf U}_n$, ${\bf A}_m^{ab}$, ${\bf B}_m^{ij}$, ${\bf G}_n$, to which we assign the conformal weights $\ell$, $\ell+1$, $\ell+1$, $\ell+2$, respectively, and their fermionic partners
${\bf S}_r^{a i}$, ${\bf R}_r^{a i}$, which have the conformal weights $\ell+\frac 12$ and $\ell+\frac 32$
\bea\label{bosbosGM}
&& \left[ L_n, {\bf U}_m\right] = -\im\, \left( \ell\, n -m\right) {\bf U}_{n+m} \;
\left[ L_n, {\bf S}_r^{a i}\right] = -\im \, \left( \left(\ell+\frac{1}{2}\right)n -r\right){\bf S}_{n+r}^{a i},\nn \\
&& \left[ L_n, {\bf A}_m^{ab}\right] = -\im\, \left( \left(\ell+1\right)\, n -m\right) {\bf A}_{n+m}^{ab} \;
\left[ L_n, {\bf B}_m^{ij}\right] = -\im\, \left( \left(\ell+1\right)\, n -m\right){\bf  B}_{n+m}^{ij} , \nn \\
&& \left[ L_n, {\bf R}_r^{a i}\right] = -\im \, \left( \left(\ell+\frac{3}{2}\right)n -r\right){\bf R}_{n+r}^{a i},\;
\left[ L_n, {\bf G}_m\right] = -\im\, \left( \left(\ell+2\right)\, n -m\right) {\bf G}_{n+m}.
\eea
As usual, the conformal weights determine the index range
\bea\label{Range}
&& {\bf U}_n: n=-\ell,\ldots,\ell, \; \quad {\bf S}_r^{a i}: r= -\ell-\frac{1}{2},\ldots,\ell+\frac{1}{2}, \; \quad {\bf A}_m^{ab}: m=-\ell-1,\ldots,\ell+1, \;\nn \\
&& {\bf B}_m^{ij}: m=-\ell-1,\ldots,\ell+1, \; \quad {\bf R}_s^{a i}: s= -\ell-\frac{3}{2},\ldots,\ell+\frac{3}{2}, \; \quad {\bf G}_n: n=-\ell-2,\ldots,\ell+2.
\eea

The remaining structure relations are unambiguously fixed by taking into account the indices which belong to the spinor representations of $su(2)$--subalgebras generated by $W^{ij}$ and $V^{ab}$, the index range in the previous formula, and the Jacobi identities
\bea\label{fermGM}
&&\left[ Q_r^{a i}, {\bf U}_m\right] = \im \; {\bf S}_{r+m}^{a i}, \nn \\
&&\left\{ Q_r^{a i}, {\bf S}_s^{b j}\right\} = \left( (2 \ell +1)r -s\right) \epsilon^{ab} \epsilon^{ij} {\bf U}_{r+s}-
\frac{2+\ell+\alpha}{3+2 \ell} \epsilon^{ab} {\bf B}_{r+s}^{ij}-\frac{1+\ell-\alpha}{3+2 \ell} \epsilon^{ij} {\bf A}_{r+s}^{ab}, \nn \\
&&\left\{ Q_r^{a i}, {\bf B}_m^{jk}\right\} =\im  \left( 2(\ell +1)r -m\right)\left( \epsilon^{ij}{\bf S}_{r+m}^{a k}+ \epsilon^{ik}{\bf S}_{r+m}^{a j}\right)
-\im \frac{1+\ell-\alpha}{3+2 \ell}\left( \epsilon^{ij} {\bf R}_{r+m}^{a k}+\epsilon^{ik} {\bf R}_{r+m}^{a j}\right), \nn \\
&&\left\{ Q_r^{a i}, {\bf A}_m^{bc}\right\} =\im  \left( 2(\ell +1)r -m\right)\left( \epsilon^{ab}{\bf S}_{r+m}^{c i}+ \epsilon^{ac}{\bf S}_{r+m}^{b i}\right)
+\im \frac{2+\ell+\alpha}{3+2 \ell}\left( \epsilon^{ab} {\bf R}_{r+m}^{c i}+\epsilon^{ac} {\bf R}_{r+m}^{b i}\right), \nn \\
&&\left\{ Q_r^{a i}, {\bf R}_s^{b j}\right\} = \left( (2 \ell +3)r -s\right)\left( \epsilon^{ab} {\bf B}^{ij}_{r+s}-\epsilon^{ij}{\bf A}_{r+s}^{ab}\right)+
\epsilon^{ij} \epsilon^{ab} {\bf G}_{r+s}, \nn \\
&&\left[ Q_r^{a i}, {\bf G}_m\right] = \im \;\left( 2(\ell+2)r -m\right) {\bf R}_{r+m}^{a i}.
\eea
In contrast to the cases discussed in the previous sections, the parameter $\alpha$ remains arbitrary. As usual, we omitted the structure relations among ${\bf U}_n$, ${\bf S}_r^{a i}$, ${\bf A}_m^{ab}$, ${\bf B}_m^{ij}$, ${\bf R}_r^{a i}$, ${\bf G}_n$ and $(W^{ij},V^{ab})$, which amount to the conventional $su(2)$--transformations.

In some sense, the superalgebra (\ref{Dalg}), (\ref{bosbosGM}), (\ref{fermGM}) is universal. The first example in Sect. 2 is obtained by setting $\alpha=\ell+1$, which correlates with (\ref{alpha1}), and choosing an invariant subalgebra spanned by $({\bf U}_n,{\bf S}_{r}^{a i},{\bf B}_m^{ij})$ and the $D(2,1;\alpha)$--generators. The second example in Sect. 2 is obtained by a reduction of (\ref{bosbosGM}) and (\ref{fermGM}). It suffices to make the formal change $\tilde\ell=\ell+2$, set $\alpha=-\tilde\ell$, which is in agreement with (\ref{alpha2}), select a subalgebra generated by $({\bf G}_n,{\bf R}_{r}^{a i},{\bf B}_m^{ij})$ and discard the rest. Moreover, the superalgebra recently introduced in \cite{GM} turns out to be a particular example of (\ref{Dalg}), (\ref{bosbosGM}), (\ref{fermGM}), which arises at $\alpha=-\frac 12$.\footnote{The formal change
$\ell=\tilde\ell-2$ is needed as well. Note that in \cite{GM} the spinor indices $i,j$ with respect to the $su(2)$--subalgebra generated by $W^{ij}$ were denoted by Greek letters $\alpha,\beta=1,2$. Our spinor notations coincide with those in \cite{GM}. In particular, $\epsilon^{12}=-1$.
} The precise relations which link the generators above and those in \cite{GM} read
\begin{align}\label{dict}
&
L_{0}=\im\, D, && L_{1}=\im\, K, && L_{-1}=\im\, H,
\nonumber\\[2pt]
&
V^{11}=-\im\, I_{+}, && V^{12}=I_{3}, && V^{22}=-\im\, I_{-},
\nonumber\\[2pt]
&
W^{ij}={\left( \sigma_a \right)}^{ij} \mathcal{J}_a, && Q_{-\frac 12}^{1 i}=\bar Q^i, && Q_{-\frac 12}^{2 i}=Q^i,
\nonumber\\[2pt]
&
Q_{\frac 12}^{1 i}=-\bar S^i, && Q_{\frac 12}^{2 i}=-S^i, && {\bf U}_n=\im\, {\bf W}^{(n+\ell-2)},
\nonumber\\[2pt]
&
{\bf S}_r^{1i}=-\im\, {\bar {\bf Z}}^{(r+\ell-\frac 32) i}, && {\bf S}_r^{2i}=-\im\, {\bf Z}^{(r+\ell-\frac 32) i}, && {\bf G}_{m}=-2\im\, {\bf C}^{(m+\ell)},
\nonumber\\[2pt]
&
{\bf R}_r^{1i}=2 {\bf L}^{(r+\ell-\frac 12) i}, && {\bf R}_r^{2i}=2{\bar {\bf L}}^{(r+\ell-\frac 12) i}, && {\bf B}_n^{ij}={\bf P}^{(n+\ell-1)ij}+{\bf P}^{(n+\ell-1)ji},
\nonumber\\[2pt]
&
{\bf A}_n^{11}=-2{\bf R}^{(n+\ell-1)}, && {\bf A}_n^{12}={\bf P}_i^{(n+\ell-1)i}, && {\bf A}_n^{22}=2{\bar {\bf R}}^{(n+\ell-1)}.
\end{align}

Thus, choosing the acceleration generators to form an analog of a generic real superfield, which is a reducible representation of $D(2,1;\alpha)$,
one can avoid the constraints on the group parameter $\alpha$ revealed in the previous sections.

\section{Realizations in superspace}

Having established the structure relations of $N=4$ $\ell$--conformal Galilei superalgebras inspired by various supermultiplets of $D(2,1;\alpha)$, let us comment on their realizations in superspace. Given a supergroup $G$, a conventional means of building its realization in superspace is to properly choose a subgroup $H \subset G$ and consider the coset space $G/H$.
Left multiplication by a group element determines the action of the supergroup on the coset whose infinitesimal form is obtained via the Baker--Campbell--Hausdorff formula.

If one wishes to realize the superalgebras above in the superspace parametrized by the temporal coordinate $t$, the odd variables $\theta_{a i}$, $a=1,2$, $i=1,2$, and a set of spatial coordinates ${\bf x}$, it suffices to include into $H$ all the generators from $G$ but for $L_{-1}$, $Q^{a i}_{-\frac{1}{2}}$, and one acceleration generator which is to be appropriately chosen in each case.
For the superalgebra (\ref{bosbos}) one can select ${\bf A}_{-\ell-1}^{ij}$, while the convenient choice for (\ref{bosbos2}) is ${\bf U}_{-\ell}$. Realization of (\ref{bosbosh}) is a bit exotic as it relies upon the fermionic generator ${\bf S}_{-\ell-\frac 12 }^{a A}$ thus asking for the fermionic spatial variables ${\bf x}$. The right choice for the superalgebra (\ref{bosbosh2}) is
${\bf q}_{-\ell}^{i A}$, for (\ref{bosbosChiral}) one should take ${\bf p}_{-\ell-1}$ and ${\bar {\bf p}}_{-\ell-1}$,
while ${\bf G}_{-\ell-2}$ provides a realization of (\ref{bosbosGM}), (\ref{fermGM}). Note that if the accelerator generators carry $su(2)$ indices, so will do the spatial coordinates ${\bf x}$ associated with them.

As an illustration, let us consider the superalgebra (\ref{bosbos2}). Constructing the coset element
\begin{equation}
g= e^{i t L_{-1}} \; e^{\theta_{a i} Q^{a i}_{-\frac{1}{2}}}\; e^{i {\bf x} {\bf U}_{-\ell}},
\end{equation}
where ${\bf x}=x^A$, and $A=1,\dots,d$ is a vector index of $so(d)$, and multiplying by a group element on the left, one obtains
infinitesimal transformations from which the following generators can be obtained
\bea\label{realiz1}
&& Q^{a i}_{-\frac{1}{2}}= \frac{\partial}{\partial \theta_{a i}} + \im \theta^{a i} \frac{\partial}{\partial t},\quad {\bf U}_{-\ell} = \im \frac{\partial}{\partial {\bf x}}, \quad {\bf S}^{a i}_{-\ell+\frac{1}{2}}= \im \theta^{a i} \frac{\partial}{\partial {\bf x}}, \quad \; L_{-1} = \im \frac{\partial}{\partial t}, \quad L_0= \im t \frac{\partial}{\partial t}+\frac{\im}{2} \theta_{a i} \frac{\partial}{\partial \theta_{a i}}+
\im \ell {\bf x} \frac{\partial}{\partial {\bf x}}, \nn\\
&& L_1 = \im \left( t^2 -\frac{1}{6} \left( 1-2 \ell \right)\theta_{a i} \theta_{b j}\theta^{b i} \theta^{a j}\right) \frac{\partial}{\partial t}+ \im \left( \theta_{a i} -\frac{\im}{3} \left( 1-2 \ell \right)\theta_{b j} \theta^b_i \theta_a^j\right)\frac{\partial}{\partial \theta_{a i}} + 2 \im \ell t {\bf x} \frac{\partial}{\partial {\bf x}}.
\eea
According to (\ref{bosbos2}), the remaining acceleration generators are obtained by computing the commutators among ${\bf U}_{-\ell}$, ${\bf S}^{a i}_{-\ell+\frac{1}{2}}$ and $L_{1}$. In particular, one can readily compute the two next members of the ${\bf U}_m$--, and ${\bf S}_r^{a i}$--towers
\bea
&& {\bf U}_{-\ell+1}=\im t \frac{\partial}{\partial {\bf x}}, \; \quad {\bf S}^{a i}_{-\ell+\frac{3}{2}}= \im \left( t \theta^{a i} -\frac{i}{3} \theta_{b j} \theta^{b i} \theta^{a j}\right)\frac{\partial}{\partial {\bf x}}.
\eea

At this point one can observe the qualitative difference with the approach in \cite{GM}. While ${\bf U}_{-\ell}$ and ${\bf U}_{-\ell+1}$ correspond to the conventional spatial translations and Galilei boosts, other members of the set ${\bf U}_{-\ell+1},\dots,{\bf U}_{\ell}$ explicitly involve the fermionic variables $\theta_{a i}$ and differ
from the conventional constant accelerations: $\im t^n \frac{\partial}{\partial {\bf x}}$, $n>2$. To put it in other words,
in this work the bosonic and fermionic acceleration generators are allowed to be polynomials in $t$ and $\theta_{a i}$, while in \cite{GM} they were chosen to have a monomial structure. Thus
the price paid for the relaxation of the constraint on the group parameter $\alpha$ in the previous section is the nonstandard form of the constant accelerations.

\section{Conclusion}

To summarize, in this work various $N=4$ supersymmetric extensions of the $\ell$--conformal Galilei algebra were constructed by properly extending the Lie superalgebra associated with the most general $N=4$ superconformal group in one dimension $D(2,1;\alpha)$. If the acceleration generators in the superalgebra formed analogues of the irreducible $(1,4,3)$--, $(2,4,2)$--, $(3,4,1)$--, and $(4,4,0)$--supermultiplets of $D(2,1;\alpha)$, the group parameter $\alpha$ was shown to be constrained by the Jacobi identities. In contrast, if the tower of the acceleration generators resembled a component decomposition of a generic real superfield, $\alpha$ remained arbitrary. It was demonstrated that an $N=4$ $\ell$--conformal Galilei superalgebra recently proposed in \cite{GM} is a particular instance of that in Sect. 5. The reason is that in \cite{GM} both the bosonic and fermionic
acceleration generators were chosen to be monomials in the temporal and fermionic coordinates, while the consideration in this work allows for a more general polynomial structure.

Turning to possible further developments, realizations of all the superalgebras in this work in terms of differential operators in superspace are of interest.
Dynamical realizations in mechanics and field theory are worth studying as well. As far as models in nonrelativistic spacetime with cosmological constant are concerned, the Newton--Hooke counterparts of the superalgebras in this work are worth studying. The structure of admissible central extensions deserves a separate consideration as well.

\vspace{0.5cm}

\noindent{\bf Acknowledgements}\\

\noindent
We thank I. Masterov for the collaboration at an earlier stage of this work and S. Fedoruk for the useful discussions.  A.G.
was supported by the Tomsk Polytechnic University competitiveness enhancement program. S.K. acknowledges the support
of the Russian Science Foundation, project No 14-11-00598.

\end{document}